\documentclass[preprints,article,accept,moreauthors,Latex,dvi2pdf,10pt,a4paper]{Definitions/mdpi}
\firstpage{1} 
\pubvolume{xx}
\issuenum{1}
\articlenumber{5}
\pubyear{2018}
\copyrightyear{2018}
\history{Submitted: September 26, 2018}
\Title{Multiplicity Dependence of the Jet Structures in pp Collisions at LHC Energies$^{\dagger}$}

\Author{Zoltán Varga $^{1,2,}$*\orcidA{}, Róbert Vértesi $^{1}$\orcidB{} and Gergely Gábor Barnaföldi $^{1}$\orcidC{}}

\AuthorNames{Zoltán Varga, Róbert Vértesi and Gergely Gábor Barnaföldi}

\address{%
$^{1}$ \quad Wigner Research Centre for Physics, P.O. Box 49, H-1525 Budapest, Hungary \\
$^{2}$ \quad Eötvös Loránd University, Egyetem tér 1-3, H-1053 Budapest, Hungary}

\corres{Correspondence: varga.zoltan@wigner.mta.hu}

\firstnote{Presented at Hot Quarks 2018 -- Workshop for young scientists on the physics of ultrarelativistic nucleus-nucleus collisions, Texel, The Netherlands, September 7-14 2018}
\abstract{We study the event multiplicity dependence of the jet structure in pp collisions. 
We present evidence for jet shape modification due to multi-parton interactions using {\sc Pythia} and {\sc Hijing++} Monte Carlo (MC) event generators as an input to our analysis. We introduce a characteristic jet size measure which is independent of the choice of the simulation parameters, parton distribution functions, jet reconstruction algorithms and even of the presence or absence of multi-parton interactions. We also investigate heavy-flavor jets and show the sensitivity of the multiplicity-differential jet structure to flavor-dependent fragmentation.
}


\begin{document}

\section{Introduction}
The collective behavior found in large systems created in heavy-ion collisions, such as long range correlations and a sizeable azimuthal anisotropy have traditionally been considered as proof for the creation of the quark-gluon plasma (QGP) in such systems.
One of the surprises in the recent years was the discovery of collective behavior in small systems, e.g. in high-multiplicity pp or pA collisions~\cite{Nagle:2018nvi}. While the presence of the QGP in such systems is still an open question, we now know that the presence of the QGP is not necessary for the explanation of collectivity, as relatively soft vacuum QCD effects such as multiple-parton interactions (MPI) can produce similar behavior~\cite{kpz}.

Jet quenching, a key signature of the QGP in AA collisions is not expected in small systems due to the insufficent volume, but in principle, effects like MPI may also cause modification of the jets in high-multiplicity events. This paper is a continuation of our previous study~\cite{VVB} aimed at the evolution of jet structure patterns and their dependence on simulation components.
 
\section{Analysis and results}

We simulated events using the {\sc Pythia} 8.226 Monte Carlo event generator~\cite{zea}. To crosscheck the results we also generated events with {\sc Hijing++}~\cite{jiz,hijproc}, a new event generator based on the still-widely-used {\sc Hijing}. We investigated three different tunes: the Monash 2013 and Monash* tunes that both use the NNPDF2.3LO PDF set~\cite{pea1,pea2}, and tune 4C with the CTEQ6L1 PDF set~\cite{ms,vw}. We simulated events with MPI switched on and off and used different Color Reconnection (CR) schemes to individually investigate the effects of these physical components on the jet structures~\cite{VVB}.
A full jet reconstruction, including both neutral and charged particles, was carried out with a resolution parameter $R=0.7$ using the anti-k$_\mathrm{T}$ algorithm of the FASTJET package~\cite{ma}. We also used the k$_\mathrm{T}$ and the Camridge-Aachen algorithms for cross-checks.
To look for jet shape modification we analyzed the transverse momentum ($p_\mathrm{T}$) distribution inside the jet cones.
We computed the  differential jet shape ($\rho$), that is the radial transverse momentum distribution inside the jet cone, as well as the integral jet shape ($\psi$), representing the average fraction of the jet transverse momentum contained inside a cone of radius $r$ around the jet axis. They are defined as

\vspace{-3mm}
\begin{eqnarray}
\rho(r) = \frac{1}{\delta r} \frac{1}{p_{\mathrm{T}}^{jet}}\sum\limits_{\substack{r_a < r_i < r_b}} p_{\mathrm{T}}^i 
& \mathrm{and} &
\psi(r) = \frac{1}{p_{\mathrm{T}}^{jet}}\sum\limits_{\substack{r_i < r}} p_{\mathrm{T}}^i
\end{eqnarray}

\noindent 
respectively, where $p_{\mathrm{T}}^i$ is the transverse momentum of a particle inside a $\delta r$ wide annulus with inner radius $r_a = r - \frac{\delta r}{2}$ and outer radius $r_b = r + \frac{\delta r}{2}$ around the jet axis and $p_{\mathrm{T}}^{jet}$ is the transverse momentum of the whole jet. The distance of a given particle from the jet axis is given by $r_i = \sqrt{(\varphi_i - \varphi_{jet})^2 + (\eta_i - \eta_{jet})^2}$, where $\varphi$ is the azimuthal angle and $\eta$ is the pseudorapidity.

\begin{figure}[H]
\centering
\includegraphics[width=0.4\linewidth]{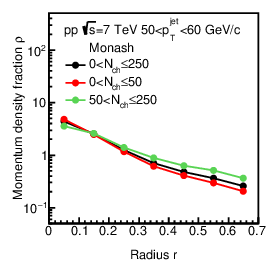}
\includegraphics[width=0.4\linewidth]{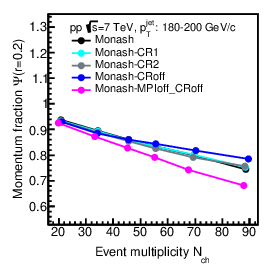}
\caption{\label{fig:rho_psi}(Left) The differential jet structure for low and high multiplicites. (Right) The integral jet structure at $r=0.2$ plotted against $N_{ch}$ for different MPI and CR simulation settings~\cite{VVB}.}
\end{figure}

\begin{figure}[H]
%\centering
\hspace{-2mm}\includegraphics[width=0.34\linewidth]{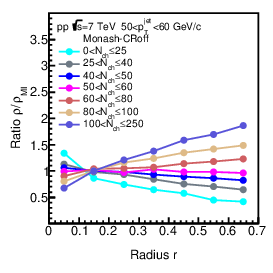}%
\includegraphics[width=0.34\linewidth]{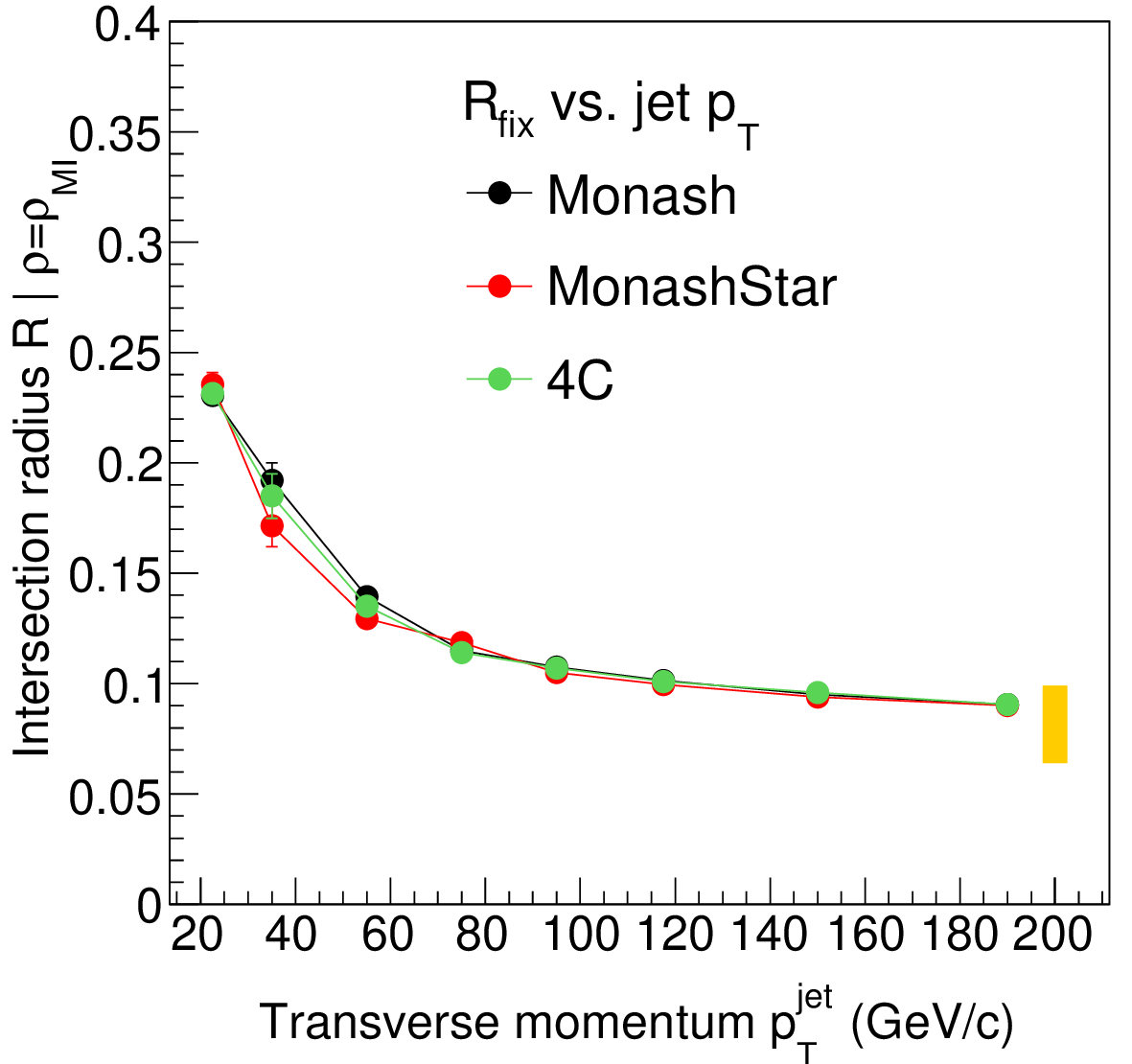}%
\includegraphics[width=0.34\linewidth]{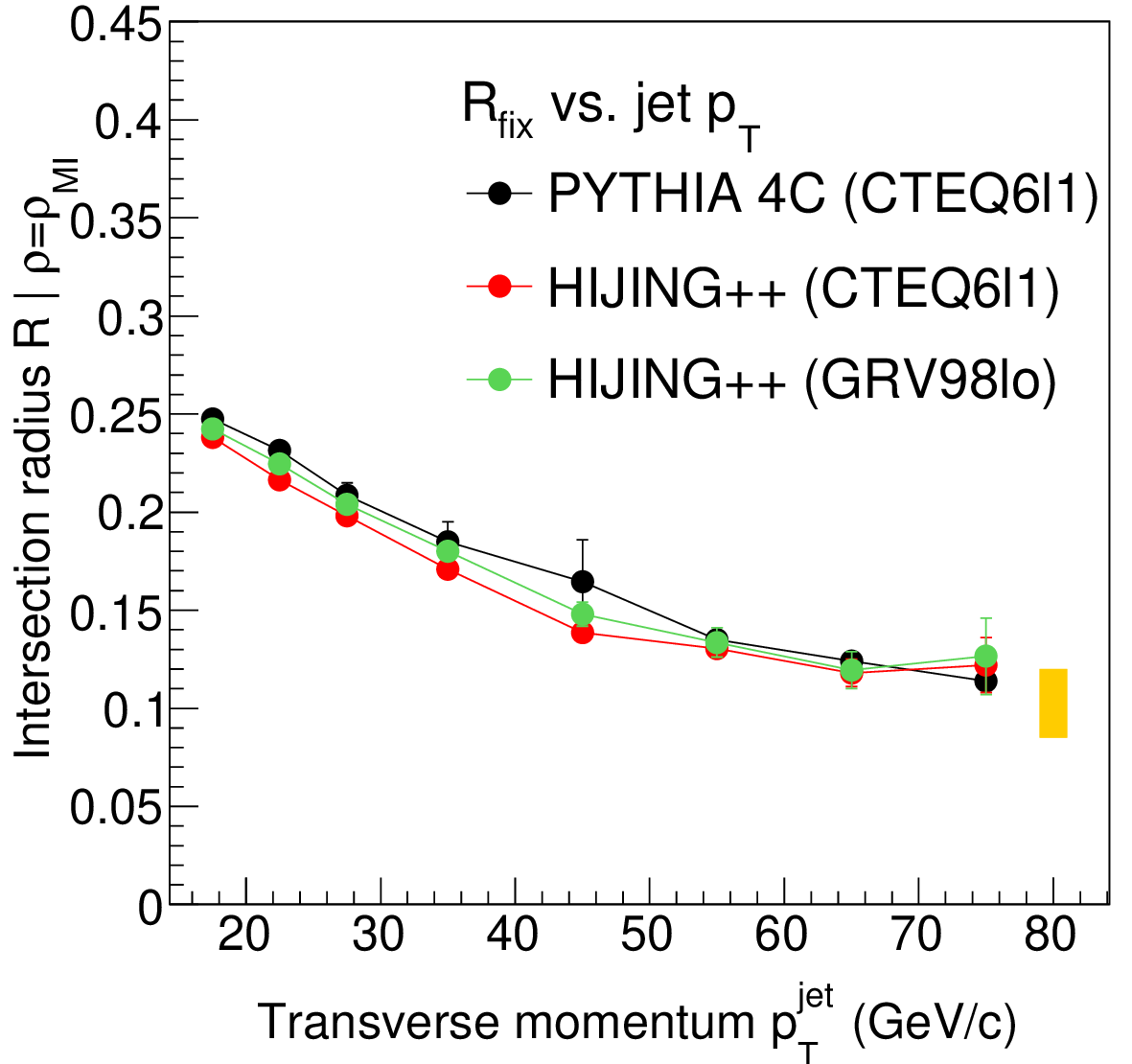}%
\caption{\label{fig:rfix}(Left) The differential jet structure compared for many different mutiplicity bins. (Center) The $p_{\mathrm{T}}^{jet}$ dependence of $R_{fix}$ for three different {\sc Pythia} tunes. (Right) Comparing the $p_{\mathrm{T}}^{jet}$ dependence of $R_{fix}$ from {\sc Hijing++} to that of {\sc Pythia} using two different PDF sets. The yellow bar indicates the systematic uncertainty stemming from the choice of $\delta r$.}
\end{figure}

Experimental data from CMS for multiplicity-integrated differential jet shapes (i.e. where events of all charged hadron multiplicities are considered) are well described by all three tunes over a wide $p_{\mathrm{T}}^{jet}$ range~\cite{Chatrchyan:2012mec,VVB}. The multiplicity-dependent analysis of the differential jet shapes, however, serve as a sensitive tool to validate simulations, as there are significant differences between predictions using otherwise equally well-preforming tunes~\cite{VVB}.
In the left panel of Figure~\ref{fig:rho_psi} we plot the differential jet structure for low and high charged hadron multiplicity ($N_{ch}$) events separately, and compare them to the multiplicity-integrated curve at a given $p_{\mathrm{T}}^{jet}$ window.
Jets in low-multiplicity events are typically more collimated than in high-multiplicity events, hence the curves intersect each other. The right panel of Figure~\ref{fig:rho_psi} shows the integral jet structure calculated at $r = 0.2$ with respect to the multiplicity. We can observe a significant effect of the MPI on the jet structures: at higher multiplicities, jets are typically less collimated if MPI is turned off.
In the left panel of Figure~\ref{fig:rfix} we plot the ratio of differential jet structures for several $N_{ch}$ classes over the multiplicity-integrated jet structure. We observe a common intersection point for all the curves. This suggests a multiplicity-independent characteristic jet-size ($R_{fix}$). Carrying out the same analysis for different tunes, settings and jet reconstruction algorithms we conclude that $R_{fix}$ depends only on the chosen $p_{\mathrm{T}}^{jet}$ window~\cite{VVB}. In the central panel of Figure~\ref{fig:rfix} we show the evolution of $R_{fix}$ with respect to the transverse momentum of the jets for different tunes. To make sure $R_{fix}$ is not simply an artefact of {\sc Pythia}, we conducted crosschecks using events generated by {\sc Hijing++}. In the right panel we show that there is no difference between the two event generators and there is also no difference between the different PDF sets~\cite{jiz,vw} we used.

\begin{figure}[H]
\centering
\includegraphics[width=0.4\linewidth]{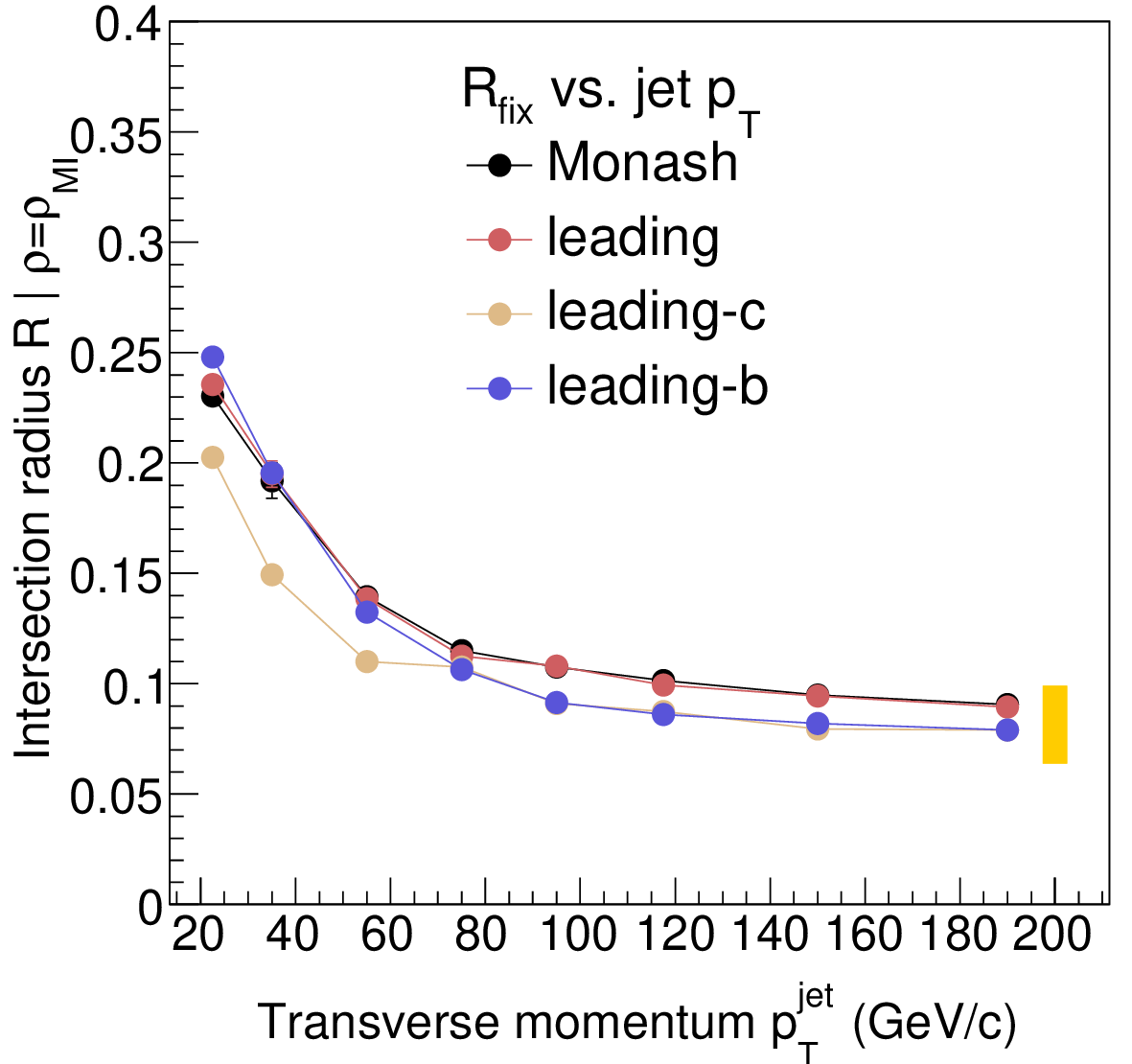}
\includegraphics[width=0.4\linewidth]{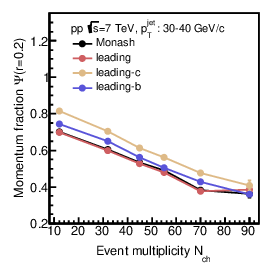}
\caption{\label{fig:hf}(Left)The $p_{\mathrm{T}}^{jet}$ dependence of $R_{fix}$ for heavy-flavor and inclusive jets. (Right) The integral jet structure at $r=0.2$ for heavy-flavor and inclusive jets.}
\end{figure}

Heavy-flavor jets are expected to undergo different fragmentation than jets from light partons, due to their different masses and color charges. Therefore we investigated the jet structures of heavy flavor jets.
In the left panel of Figure~\ref{fig:hf} we compare the $R_{fix}$ for selected leading and subleading jets with the default configuration where any jet is allowed (labeled as Monash in the figure). Selecting only the leading and subleading jets from events does not make a significant difference. However, the $R_{fix}$ of leading and subleading jets with beauty or charm content shows slightly different patterns. At low $p_{\mathrm{T}}^{jet}$, c-jets have slightly lower $R_{fix}$ than inclusive and b-jets, while at high $p_{\mathrm{T}}^{jet}$, heavy-flavor jets differ from inclusive jets.
In the right panel we show the integral jet structure at $r = 0.2$ for heavy-flavor jets. Here we can observe that heavy-flavor jets are narrower than inclusive jets on the average.
This clearly shows that multiplicity-differential jet structures are sensitive to flavor-dependent fragmentation.
However, the effect is not ordered by mass, since the beauty  $\psi(r=0.2,N_{ch})$ curve is between the inclusive and the light jets. It is also to be noted that the ordering is $p_{\mathrm{T}}^{jet}$-dependent, with a similar trend to the one observed in $R_{fix}$. 

\newpage
%%%%%%%%%%%%%%%%%%%%%%%%%%%%%%%%%%%%%%%%%%
\section{Conclusions}

A systematic study on the charged hadron event multplicity dependence of the jet structures has been carried out using {\sc Pythia} and {\sc Hijing++} MC event generators. We showed that the multi-parton interactions cause a significant modification of the jet structure. 
We have proposed a multiplicity-independent characteristic jet size measure, $R_{fix}$, which is independent of the choice of simulation parameters, parton distribution functions, jet reconstruction algorithms and even of the presence or absence of multi-parton interactions. There is no significant difference between $R_{fix}$ obtained by {\sc Pythia} and {\sc Hijing++}.
Its $p_{\mathrm{T}}^{jet}$ dependence can be qualitatively explained by the Lorentz-boost of the jet~\cite{VVB}.
Finally, we showed that multiplicity-differential jet structure measurements provide a sensitive probe of flavor-dependent fragmentation.
Comparing the jet structures observed in heavy-flavor jets and inclusive jets shows a
$p_{\mathrm{T}}^{jet}$-dependent difference, which, however, does not follow mass ordering.

%%%%%%%%%%%%%%%%%%%%%%%%%%%%%%%%%%%%%%%%%%
\vspace{6pt} 

%%%%%%%%%%%%%%%%%%%%%%%%%%%%%%%%%%%%%%%%%%
\authorcontributions{Software and formal analysis, Z.V., R.V.; investigation, R.V., Z.V., G.G.B.; writing—original draft preparation, Z.V.; writing—review and editing, R.V.; supervison, R.V.; funding acquisition, G.G.B.}
%\authorcontribution{Software, Z.V. and R.V.; formal analysis, Z.V., R.V and G.G.B.; investigation, Z.V., R.V. and G.G.B.; writing—original draft preparation, Z.V.; writing—review and editing, Z.V. and R.V.; supervision, Z.V..}

%%%%%%%%%%%%%%%%%%%%%%%%%%%%%%%%%%%%%%%%%%
\funding{This work has been supported by the NKFIH/OTKA K 120660 grant, the János Bolyai scholarship of the Hungarian Academy of Sciences (R.V.) and the MOST-MTA Chinese-Hungarian Research Collaboration. The work has been performed in the framework of COST Action CA15213 THOR.}

%%%%%%%%%%%%%%%%%%%%%%%%%%%%%%%%%%%%%%%%%%
\acknowledgments{The authors would like to thank for the many useful conversations they had with Jana Bielčíková, Gábor Bíró, Miklós Kovács, Yaxian Mao and Gábor Papp.}

%%%%%%%%%%%%%%%%%%%%%%%%%%%%%%%%%%%%%%%%%%
\conflictsofinterest{The authors declare no conflict of interest.} 

%=====================================
% References, variant A: internal bibliography
%=====================================
\reftitle{References}

%%%%%%%%%%%%%%%%%%%%%%%%%%%%%%%%%%%%%%%%%%
\end{document}